\documentclass{epl}

\title{Observation of a Transient Magnetization Plateau in a Quantum Antiferromagnet
on the Kagom\'{e} Lattice}

\author{Y. Narumi\inst{1}\thanks{Present address: KYOKUGEN, Osaka University,
Machikaneyama, Toyonaka, Osaka 560-8531, Japan.}\thanks{E-mail:
\email{narumi@mag.rcem.osaka-u.ac.jp}} \and K. Katsumata\inst{1} \and
Z. Honda\inst{2} \and J.-C. Domenge\inst{3} \and P. Sindzingre\inst{3}
\and C. Lhuillier\inst{3} \and Y. Shimaoka\inst{4} \and T. C. Kobayashi\inst{4}
\and K. Kindo\inst{4}}

\institute{
\inst{1} RIKEN Harima Institute, Mikazuki, Sayo, Hyogo 679-5148, Japan \\
\inst{2} Faculty of Engineering, Saitama University, Shimo-Okubo, Saitama338-8570,
Japan \\
\inst{3} Laboratoire de Physique Th\'{e}orique des Liquides, Universit\'{e} P. et
M. Curie and UMR 7600 of CNRS, case 121, 4 Place Jussieu, 75252 Paris Cedex, France\\
\inst{4} KYOKUGEN, Osaka University, Toyonaka, Osaka 560-8531, Japan
}


\begin{document}

\maketitle

\begin{abstract}
The magnetization process of an $S=\frac{1}{2}$ antiferromagnet on the kagom\'{e}
lattice, [Cu$_{3}$(titmb)$_{2}$(OCOCH$_{3}$)$_{6}$]$\cdot$H$_{2}$O\{titmb=
1,3,5-tris(imidazol-1-ylmethyl)-2,4,6 trimethylbenzene\} has been measured at very
low temperatures in both pulsed and steady fields.  We have found a new dynamical
behavior in the magnetization process.  A plateau at one third of the saturation
magnetization $M_{\rm s}$ appears in the pulsed field experiments for intermediate
sweep rates of the magnetic field and disappears in the steady field experiments.
A theoretical analysis using  exact diagonalization yields, $J_1 = -19 \pm 2 K$ and
$J_2 = 6 \pm 2  K$, for the nearest neighbor and second nearest neighbor interactions,
respectively.  This set of exchange parameters  explains the very low saturation field
and the absence of the plateau in the thermodynamic equilibrium as well as the two-peak
feature in the magnetic heat capacity observed by Honda et al. [ Z. Honda {\it et al.},
J. Phys.: Condens. Matter {\bf 14}, L625 (2002)].  Supported by numerical results we
argue that a dynamical order by disorder phenomenon could explain the transient
appearance of the  $M_{\rm s}/3$ plateau in pulsed field experiments.
\end{abstract}

The study of quantum phenomena in magnetic systems has been a central subject in
condensed matter research in recent years.  Lower dimensional antiferromagnets show
many interesting phenomena originating from quantum fluctuations.  One of the examples
of these phenomena is the appearance of a singlet ground state and a spin gap to the
excited triplet in a spin, $S$=1 one-dimensional Heisenberg antiferromagnet (1D HAF)
\cite{haldane}, a spin-Peierls system\cite{bray} and an $S$=$\frac{1}{2}$ two-leg spin
ladder\cite{dagotto}.  Another interesting example is the appearance of a magnetization
plateau, a phenomenon in which the magnetization, $M$, of an antiferromagnet stays
constant in a finite range of applied magnetic fields, $B$, before reaching the
saturated phase.  In the last few years, magnetization plateaus have been observed in
several quantum antiferromagnets including an $S$=1 1D antiferromagnet with bond
alternation\cite{narumi98}, the coupled dimer compound NH$_{4}$CuCl$_{3}$\cite{tanaka98}
and SrCu$_{2}$(BO$_{3}$)$_{2}$\cite{kageyama99,miyahara99,muller00,totsuka01,misguich01,kodama02}.
The origin of the magnetization plateau in these quantum antiferromagnets differs for
different compounds since a magnetization plateau can be expected to occur from various
combinations of dimensionality and frustration of interactions\cite{claire02}.  Theory
predicted that a quantum antiferromagnet on the kagom\'{e} lattice with a nearest
neighbor Heisenberg interaction, which is a typical example of frustrated magnets,
would also show a magnetization plateau at one third of the saturation magnetization
$M_{\rm s}$\cite{claire02,hida01,cabra02}.  There has been, however, no experimental
finding of the magnetization plateau in kagom\'{e} antiferromagnets.  In this paper,
we report the first observation of a magnetization plateau in the $S$=$\frac{1}{2}$
kagom\'{e} antiferromagnet, [Cu$_{3}$(titmb)$_{2}$(OCOCH$_{3}$)$_{6}$]$\cdot$H$_{2}$O,
\{titmb=1,3,5-tris(imidazol-1-ylmethyl)-2,4,6 trimethylbenzene\} (hereafter, abbreviated
to Cu-titmb).

First, we summarize the crystal and magnetic properties of Cu-titmb.  The crystal
structure of this material has been determined by Liu $et$ $al.$ \cite{liu99} and is
shown schematically in Fig. 1.  As is seen from Fig. 1(a), Cu$^{2+}$ ions are placed
at the vertices of the triangles.  The distance between the Cu$^{2+}$ ions on a triangle
is 7.77 \AA.  These triangles are attached to a hexagon.  A two dimensional array of
such pattern is called the kagom\'{e} lattice.  The kagom\'{e} layers are stacked
along the $c$ axis as shown in Fig. 1(b) with the inter layer distance of 7.05 \AA.
We define the nearest neighbor exchange interaction, $J_{1}$, between Cu$^{2+}$ spins
in a kagom\'{e} layer as the one along the edges of a triangle through H$_{2}$O and/or
CH$_{3}$COO$^{-}$.  The former path is expected to dominate because the latter path
is connected by a weak hydrogen bond.  The second neighbor interaction, $J_{2}$, is
defined as the one between a spin and the spins on the second neighboring vertices of
the hexagon.  Since no apparent exchange paths exist between spins on the adjacent
kagom\'{e} layers along the $c$ axis, the interaction between layers is expected to
be much smaller than $J_{1}$ and $J_{2}$.

Heat capacity measurements were performed on a powder sample of Cu-titmb\cite{honda02}.
The magnetic heat capacity, $C_{\rm m}$, obtained after subtraction of the lattice heat
capacity is shown in the inset of Fig. 2 as a function of temperature, $T$.  We see a
two peak structure which is a feature predicted for an $S$ = $\frac{1}{2}$ aniferromagnet
on the kagom\'{e} lattice with the nearest neighbor Heisenberg
interaction\cite{elstner94,nakamura95,sind00} and for other frustrated quantum
magnets\cite{claire02}.

Powder samples of Cu-titmb were grown by the method described in\cite{liu99}.  The
material, 1,3,5-tris(imidazol-1-ylmethyl)-2,4,6-trimethylbenzene, was purchased from
the Wako Pure Chemical Industries, Ltd.  The magnetization measurements were performed
at Osaka University under pulsed high fields in conjunction with a dilution refrigerator.
Pulsed magnetic fields are generated by discharging the energy stored in a capacitor
bank through the coil.  The duration time of the pulsed field is given roughly by
$(LC)^{\frac{1}{2}}$, where $L$ is the inductance of the coil and $C$ is the capacitance
of the bank.  Since the magnetic field changes sinusoidally with time, the field sweep
rate, $dB/dt$, differs for different field values.  In the following, we denote the
field sweep rate as the one obtained at around 1.5 T.  To change the sweep rate, we
have to change either, $C$, $L$ and/or the peak field.  In this experiment, we used
two sets of capacitor banks and were able to measure the magnetization over a wide
range of sweep rates.  However, we were not able to access the sweep rates below
0.7 T/msec and those between 1.9 T/msec and 14 T/msec from the technical reason
described above.  The powder sample was packed into a Teflon tube and immersed in the
mixing chamber of the dilution refrigerator.  In the pulsed field measurements, we
observe the derivative of magnetization with respect to time, $dM/dt$.  Simultaneously,
we measure the time evolution of $B$.  Combining these two sets of data, we get $M$ vs
$B$ curve.  Figure 2 shows the magnetization as a function of applied magnetic field
measured at 100 mK in increasing and  decreasing fields.  We see a magnetization
plateau at $M_{\rm s}$/3 for 0.5 T $\leq B \leq$ 2.5 T in the increasing field run.
Then the magnetization increases almost linearly with increasing field up to saturation.
The magnetization becomes saturated above about 10 T with the saturation moment of
about 1 $\mu_{\rm B}$ per Cu atom.  This value of $M_{\rm s}$ is consistent with the
value expected for a Cu$^{2+}$ with $S$ = $\frac{1}{2}$ and $g$ = 2.20\cite{liu99}.
Surprisingly, no plateau has been observed in decreasing field. Figure 3 shows the
magnetization curve measured in increasing fields with different sweep rates.  As the
field is swept slower, the $M_{\rm s}$/3 plateau becomes less pronounced.  When
$B$ is swept much faster, we do not see the plateau either.  The results presented in
Figs. 2 and 3 suggest that the $M_{\rm s}$/3 plateau might be metastable.  In order
to get a magnetization curve in equilibrium, we have measured the magnetization of
Cu-titmb by the Faraday method in steady fields using a superconducting
magnet\cite{koba01}.  A powder sample of Cu-titmb was mixed with Apiezon-N grease and
put on one of the plates of the capacitor.  In order to ensure a good thermometry, Cu
foils were used to link the sample and the mixing chamber.  Figure 4 shows the results.
In the thermodynamic limit, the magnetization increases abruptly with increasing
fields from almost zero and no plateau is seen.  Saturation of the magnetization is
observed at $B_{\rm sat}\sim 0.2$ T.

In the following, we discuss theoretically the experimental results reported above.
The proximity of a ferromagnetic instability due to a competition between ferromagnetic
and antiferromagnetic exchange interactions  is the simplest explanation for the
anomalously low saturation field, $B_{\rm sat}$, measured in the static experiments.
Such  competing interactions may be described by a $J_1-J_2$ model on the kagom\'e
lattice, with $J_1\times J_2 <0$.  The model reads:

\begin{equation}
{\cal H} =  J_1 \sum_{<i,j>} {\bf S}_i.{\bf S}_j + J_2 \sum_{<<i,k>>} {\bf S}_i.{\bf S}_k
\end{equation}
where the first and second sums run on the first and second neighbors respectively.
For pure antiferromagnetic $J_{1}$ (or $J_{2}$) couplings and in a finite range of
parameters around these situations the system is in the special spin liquid phase
which has been the subject of many studies\cite{elstner94,nakamura95,waldtmann98}.
A large enough second neighbor ferromagnetic interaction drives the system towards
the so-called $\sqrt{3} \times \sqrt{3}$ three-sublattice N\'eel order, whereas a
combination of antiferromagnetic $J_{1}$ and $J_{2}$ induces a $q$ = 0 three-sublattice
N\'eel order\cite{lblps97}.  A ferromagnetic phase exists over a large range of $J_{1}$
and $J_{2}$, which become unstable when the antiferromagnetic second-neighbor
interaction becomes large ($J_{2}/J_{1} > - 0.35$).  As will be explained below, we
found that Cu-titmb is in this range of parameters very close to the ferromagnetic
phase.  For these coupling parameters a numerical calculation up to 36 sites gives
the quantum eigen-spectra which show a 12 sublattice N\'eel order with a non coplanar
order parameter, in agreement with the classical solution. A spin wave calculation
remains to be done to give a plausible estimate of the reduction of the order
parameter by long-wave-length quantum fluctuations\cite{dsl03}.  In the experiments
$B_{\rm sat}$ is very low compared to $T_{\rm max}$, the temperature at which
$C_{\rm m}$ shows a broad peak (see Fig. 2): the dimensionless parameter,
$\theta =(g\mu_{\rm B} B_{\rm sat})/(k_{\rm B} T_{\rm max})$ is approximately 0.02,
whereas this ratio is expected to be of the order of one in an unfrustrated
antiferromagnet and is 4.5 in the pure antiferromagnetic model on the
kagom\'e lattice\cite{claire02,hida01,cabra02,elstner94}.  In the $J_1-J_2$ model,
$\theta$ decreases when approaching the boundary of the ferromagnetic phase\cite{dsl03}.
A value of $\theta$ consistent with the experimental one is obtained in the vicinity
of the ferromagnetic instability either with ferromagnetic $J_1$ and antiferromagnetic
$J_2$ with $J_2/J_1 \sim -0.3$, or with antiferromagnetic $J_1$ and ferromagnetic
$J_2$ with $J_2/J_1 \sim -50$.  Mizuno $et$ $al.$\cite{mizuno} have calculated the
exchange interactions in copper oxides.  They showed that the exchange interaction
between Cu$^{2+}$ spins through an oxygen atom is ferromagnetic when the bond angle is
around 90$^{\circ}$.  We, therefore, adopt a ferromagnetic $J_{1}$. 

A reasonable agreement between the experimental data for $T_{\rm max}$ and $B_{\rm sat}$
is obtained with $J_1 = -19 \pm 2 K$ and $J_2 = 6 \pm 2  K$.  This set of $J_{1}$ and
$J_{2}$ also predicts a double peak feature in $C_{\rm m}$ as observed experimentally
(Fig. 2).  At the thermodynamic limit, this model has no magnetization plateau as will
be shown below.

Such a competition of interactions induces a very high density of low lying magnetic
states.  Numerous magnetic excitations are thus at very low energy and very small
perturbations can change the magnetic behavior.  We have observed in our numerical
study a general phenomenon that might be relevant to the real physical behavior of the
compound.  Samples with $N$= 21 and 27 which do not accommodate the wave vectors at
(0, $\pi$), ($\pi$, 0) or ($\pi$, $\pi$) in their Brillouin zone frustrate the
thermodynamic ground state and display a $\frac{1}{3}$ plateau as shown in Fig. 5.
In this figure, is shown the dependence of the field range, $B_{2}-B_{1}$, at which
the magnetization plateau at $m(\equiv M/M_{s})=\frac{1}{3}$ appears on the energy per
spin, $e(m)$, in this state.  At the thermodynamic limit shown schematically in the
inset, a cusp in the e(m) curve at m=m$_{\rm 0}$ is the signature of a m$_{\rm 0}$
magnetization plateau bounded by the two fields $B_1$ and $B_2$ which measure the
slopes of the e(m) curve on the left and right of $m_{\rm 0}$\cite{claire02}. Exact
diagonalization on finite size samples gives the energy per spin versus magnetization
for the discrete values m=2S$_{\rm tot}$/N. These energies are then fitted to two
polynomial forms allowing an eventual discontinuity of the slopes at the specific point
m$_{\rm 0}$. The best fits are obtained with a discontinuity at $m=\frac{1}{3}$ and
give the width $B_2 - B_1$ of the magnetization plateau reported in the figure. The
error bars represent the uncertainties in the parameters of each fit. The results for
the sizes multiples of 6 (N=18, 24, 30 and 36) exhibit zero width ($i. e.$ no plateau)
in the thermodynamic limit. The N=21 and 27 samples, which do not allow the wave
vectors (0,$\pi$), ($\pi$,0), ($\pi$,$\pi$) have a higher energy per spin and exhibit
a plateau at $m=\frac{1}{3}$.  At any magnetization the energy per spin of these
samples at $T$=0 is indeed slightly higher than the true thermodynamic ground-state
and thus the system is in a metastable configuration.  At $M_{\rm s}$/3, this metastable
state is associated with an up-up-down ($uud$) configuration of spins, where spins
along the horizontal lines in Fig. 1(a) are ferromagnetically aligned in the up
direction, whereas the extra spins of the triangles are down\cite{claire02}.  Three
such configurations are possible and thus we have three domains.

Among all the configurations of spins that are at very low energy at $M_{\rm s}$/3,
purely geometric arguments imply that the three $uud$ configurations have in phase
space the largest attraction basin for fluctuations, either thermal or quantum, and
thus a lock-in of the system in such configurations can be driven by a large number of
very small perturbations.  This is a transient analogue of the phenomenon of order
by disorder\cite{villain80,shend82,henley89}.  In such a scenario there exist in the
free energy landscape three large metastable wells associated with the three possible
$uud$ configurations.  Two characteristic times of the out of equilibrium dynamics in
such a landscape are then of importance: the relaxation time, $\tau_{\rm r}$, which is
dominated by the curvature of the metastable well, and the Kramers escape time from
the well, $\tau_{\rm esc}$, which involves both the characteristics of the well and of the
surrounding energy barriers.  If the sweep rate, $\it \Gamma$, of the magnetic field
is larger than $\tau_{\rm r}^{-1}$ no signature of the metastable well can be seen.
If $\it \Gamma$ is smaller than $\tau_{\rm esc}^{-1}$, the system relaxes to the
equilibrium situation and no magnetization plateau is observed.  For sweep rates
$\tau_{\rm esc}^{-1} < \it \Gamma <  \tau^{- 1}_{\rm r}$, the system can stay for a
while in the metastable well and display a $M_{\rm s}$/3 plateau.  Such a picture of
the dynamics should be supported by a stochastic description of the evolution of a
small cluster of spins with a Fokker Planck equation.  In such an approach
typically $\tau_{r} = \gamma / \omega_{0}^{2}$ where $\omega_{0}$ is the smallest
spin wave pulsation in the metastable configuration and $\gamma$ is a phenomenological
friction coefficient.  Whatever the origin of the escape from the metastable well,
$\tau_{esc}$ is noticeably larger than $\tau_{r}$: it involves a factor which depends
exponentially on the height, $\Delta$ of the potential energy barrier.  In the
classical regime typically $\ln (\tau_{esc})$ is proportional to $\Delta/(\chi
\gamma D)$, where $\chi$ is the homogeneous susceptibility of the cluster and $D$ the
spin diffusion coefficient.  In the quantum tunnelling regime, which is more probable
in the present case,  $\ln (\tau_{esc}) \propto \sqrt(\Delta \chi)$.  More theoretical
and experimental work is indeed needed to clarify this point.  

In conclusion, we have measured the magnetization process of the $S=\frac{1}{2}$
antiferromagnet on the kagom\'{e} lattice, Cu-titmb, at very low temperatures in both
pulsed and steady fields.  We have found for the first time a $M_{\rm s}/3$ plateau in a
certain range of field sweep rates in the pulsed field measurements.  A theoretical
analysis using  exact diagonalization yields, $J_1 = -19 \pm 2 K$ and $J_2 = 6 \pm 2  K$,
for the nearest neighbor and second nearest neighbor interactions, respectively.
This set of exchange parameters also explains the two-peak feature in the magnetic
heat capacity observed by Honda $et$ $al.$\cite{honda02}.  This  very strong  frustration
and proximity of the ferromagnetic phase is the origin of a very large density of low lying
magnetic states.  Amongst these states the $uud$ configurations are not very far in
energy from the absolute ground state and can be selected by an effect of order by
disorder.  This might be an explanation of the transient appearance of a metastable
plateau at $M_{\rm s}/3$ with appropriate sweeping rates of the magnetic field. 
 
\acknowledgments
 
We would like to thank A. Matsuo and K. Suga for their help in the pulsed field
measurements.  This work was supported in part by a Grant-in-Aid for Scientific Research
from the Japan Society for the Promotion of Science. Computations were performed at the
Centre de Calcul pour la Recherche de l'Universit\'e Pierre et Marie Curie and at the
Institut de d\'eveloppement des Recherches en Informatique Scientifique du CNRS under
contract 020076.

\newpage

\begin{figure}
\caption{The crystal structure of
 [Cu$_{3}$(titmb)$_{2}$(OCOCH$_{3}$)$_{6}$]$\cdot$H$_{2}$O.}
\label{f.1}
\end{figure}

\begin{figure}
\caption{The magnetization versus applied magnetic field curve of
[Cu$_{3}$(titmb)$_{2}$(OCOCH$_{3}$)$_{6}$]$\cdot$H$_{2}$O measured
in increasing and decreasing field runs at $T$ = 100 mK.  Inset:
The temperature dependence of the magnetic heat capacity of
[Cu$_{3}$(titmb)$_{2}$(OCOCH$_{3}$)$_{6}$]$\cdot$H$_{2}$O measured
at the designated magnetic fields.}
\label{f.2}
\end{figure}

\begin{figure}
\caption{The magnetization versus applied magnetic field curve obtained
in increasing field runs with the designated sweep rates at $T$ = 100 mK.}
\label{f.3}
\end{figure}

\begin{figure}
\caption{The magnetization versus applied magnetic field curve obtained
in a steady field at the designated temperatures.}
\label{f.4}
\end{figure}

\begin{figure}
\caption{Dependence of the field range, $B_{2} - B_{1}$, at which the
$m = \frac{1}{3}$ plateau appears on the energy per spin, $e(m)$ in this
state. Inset: $e$ versus $m$ in the thermodynamic limit.}
\label{f.5}
\end{figure}

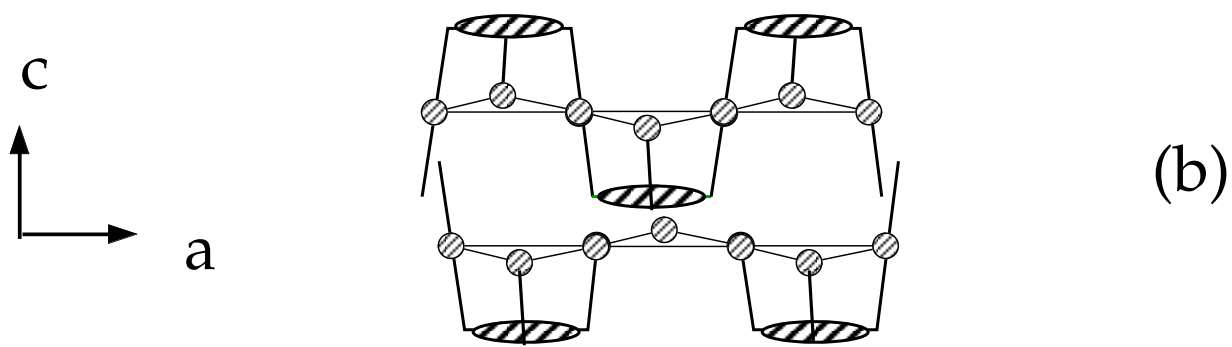
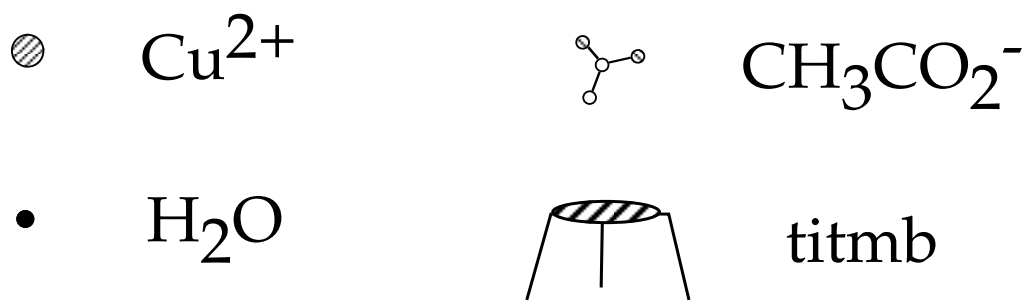
\end{document}